\documentclass[fleqn,usenatbib]{mnras}

\usepackage{newtxtext,newtxmath}

\usepackage[T1]{fontenc}

\DeclareRobustCommand{\VAN}[3]{#2}
\let\VANthebibliography\thebibliography
\def\thebibliography{\DeclareRobustCommand{\VAN}[3]{##3}\VANthebibliography}

\usepackage{orcidlink}
\usepackage{gensymb}
\usepackage{caption}
\usepackage{subcaption}
\usepackage{comment}
\usepackage{graphicx}	
\usepackage{amsmath}	

\title[The RSD Lightcone after Foreground Removal]{Recovering the Coupled Treatment of Redshift-Space Distortions and the Lightcone Effect after Diffuse Foreground Removal}

\author[J. Feron et al.]{Jennifer Feron$^{1}$\,\orcidlink{0009-0001-8626-540X}\thanks{E-mail: ppxjf3@nottingham.ac.uk},Emma Chapman$^{1}$\,\orcidlink{0000-0002-5050-9847}
\\
$^{1}$School of Physics and Astronomy, The University of Nottingham, University Park, Nottingham, NG7 2RD, UK\\
}

\date{Accepted XXX. Received YYY; in original form ZZZ}

\pubyear{\the\year{}}

\begin{document}
\label{firstpage}
\pagerange{\pageref{firstpage}--\pageref{lastpage}}
\maketitle

\begin{abstract}
The $21$ cm brightness temperature during the Epoch of Reionisation is widely modelled using semi-numeric simulations, used for their computational speed and flexibility in testing astrophysical and cosmological parameters. However, it is common practice to simulate coeval brightness temperature boxes, and then apply post-processing algorithms that treat the lightcone effect and redshift-space distortions separately, assuming they can be added in sequence. We instead model them together, allowing for partial coeval cell contributions, and ensuring that velocity-induced frequency shifts are computed at the correct cosmic time for every position along the line of sight. We show that considering these effects simultaneously creates a difference in the shape of the power spectrum over all Fourier scales, and remains recoverable after semi-blind foreground removal. We show that our lightcones consist of an average of 8\% and maximum of 120\% of a coeval cell length. These contributions to a 21cm brightness temperature lightcone voxel are shifted from within a $\pm0.5$ MHz range of the emitted frequency. The boost in the power spectrum seen over small scales ($k>1.5 $ Mpc) of our robust $21$ cm lightcone method compared to basic methods is recoverable after the addition and removal of diffuse radio foregrounds. The largest differences during the Epoch of Reionisation lie in the $k$-space, where the noise sensitivity for a 1000-hour SKAO-low observation is greater than the signal. However, in the cosmic dawn, we have shown that the major differences lie outside of this noise-dominated region.
\end{abstract}




\section{Introduction}
Neutral Hydrogen ($\rm HI$) first formed about $380,000$ years after the Big Bang, when the Universe cooled enough to allow electrons and protons to combine \citep{planck_results_2014}. This transition allowed the Universe to become transparent to photons, and small fluctuations in the temperature and density fields were imprinted on the Cosmic Microwave Background (CMB). As the Universe expanded and cooled, this background radiation shifted out of the visible wavelengths, beginning the Dark Ages. When $\rm HI$ had cooled enough such that high-density pockets could coalesce and collapse, the conditions were right for fusion processes to begin at the core of these regions. Hence, the first luminous sources were formed, ending the Dark Ages, around 180 million years after the Big Bang \citep{BarkanaLoeb2001}. The first stars emitted ultraviolet (UV) photons, while the first active galactic nuclei (AGNs) emitted  X-rays, both of which ionised the surrounding $\rm HI$ regions. This returned the $\rm HI$ to the ionised state it was in just after the Big Bang, carving out bubbles of ionised hydrogen ($\rm HII $). These bubbles expanded and merged over several hundred million years, during a period known as the Epoch of Reionisation (EoR) \citep[see e.g.][]{BarkanaLoeb2001,Furlanetto_expandHII_2004,Iliev_largescales_2006, Robertson_star_2010}. From observational constraints, such as quasar absorption spectra, Lyman $\alpha$ (Ly$\alpha$) emitters and the Thompson optical depth of CMB photons, a picture of reionisation has been built. These constraints detail that the EoR began at $z \approx 14$, and was completed by $z\approx 5-6$ \citep{Robertson_star_2010,Fan_obsconstraints_2006}, with a midpoint at $z\approx 7-8$ \citep[e.g.][]{Davies2018,Yang_opical_depth_2020,Qin2021,Nakane2023,Umeda_nhi_2023,Durovcikova2024}. A late ($z\sim5.3$), extended, patchy end to reionisation is currently favoured due to observations of both the Ly$\alpha$  forest \citep{Becker2015,eilers_opacity_2018,bosman_new_2018,bosman_hydrogen_2022} and of neutral islands persisting up to $z < 6$ \citep{fan_constraining_2006, becker_island_2024}. 

Another observable constraint on the EoR is the brightness temperature of the $21$ cm line. This line arises from the hyperfine splitting of the $1\rm S$ ground state of neutral hydrogen. This splitting is caused by the magnetic moments of the proton and electron, creating two energy levels. This so-called `forbidden transition' has a very small transition rate of $\approx10^{-15} \, \rm s^{-1}$, however, due to the large abundance of $\rm HI$ in the Universe, this line can be used to trace cosmic history. A detection of the $21$ cm line originating from the EoR remains elusive due to the complexity of the interferometers used to detect it, as well as the presence of radio foregrounds that can be up to 4-5 magnitudes larger than the predicted signal \citep[e.g.][]{Morales_ps_sensitivity_2005,McQuinn_2006}. The first step is to remove bright radio point sources during the calibration steps for each telescope. These can be removed using direction-dependent calibration and then cataloguing the remaining sources \citep{GLEAM_2017} to iteratively remove them with algorithms such as $\rm WSClean$ \citep{Offringa_CLEAN_2014,Yatawatta_calibartion_2024}. The main diffuse radio foreground emission arises from synchrotron and free-free processes inside the Milky Way, as well as extragalactic emission from AGNs and star-forming galaxies \citep{Jelic_foregrounds_2008}. This diffuse emission is removed using one of two main approaches. Foreground avoidance exploits the fact that there is a region in Fourier space theoretically free from foregrounds. This foreground-free space, known as the `EoR window', occurs due to the frequency smoothness of the foregrounds and the frequency-dependent response of the instrument used \citep[e.g.][]{Datta_brightsources_2010,Morales_foregrounds_2012, Vedantham_foregrounds_2012, Trott_foregrounds_2012, Parsons_foregrounds_2012,  Dillon_foreground_wedge_2014, Liu_wedge_2014, Pal_fgavoidance_2022}. Foreground removal aims to model and subtract the foregrounds from the signal. A variety of different removal methods exist. These include parametric, using knowledge about the foregrounds and instrument \citep{Mertens_GPR_2018}, and non-parametric \citep[e.g.][]{Liu_PCA_2011, Chapman_fastica_2012, Chapman_GMCA_2013, Switzer_SVD_2015, Zuo_SVD_2023} techniques. Any method to deal with foreground contamination requires a robust understanding of the foregrounds and instruments used, as well as how they move power across Fourier modes and contaminate the `EoR window'. Current radio arrays, such as the Low Frequency Array (LOFAR)\footnote{https://www.astron.nl/telescopes/lofar/} \citep{van_Haarlem_2013}, the upgrade to LOFAR NenuFAR \footnote{https://nenufar.obs-nancay.fr/en/homepage-en/}\citep{Zarka_NenuFAR_2012}, Murchison Widefield Array (MWA)\footnote{https://www.mwatelescope.org/} \citep{Bowman_MWA_2013} and Hydrogen Epoch of Reionisation Array (HERA)\footnote{https://reionization.org/} \citep{DeBoer_HERA_2017} are using these techniques to provide upper limits on the $21$ cm brightness temperature during the EoR \citep{Munshi_21limits_2024, HERA_limits_2023, Yoshiura_MWA_21limits_2021, mertens_improved_2020}. The Square Kilometre Array Observatory\footnote{https://www.skao.int/en} (SKAO), currently under construction, will hopefully be able to obtain images of the ionised bubbles created during the EoR and provide information on the physical conditions governing this time. This is because it provides a great increase in sensitivity compared to the current generation of radio interferometers \citep{Koopmans_2015}.

In this work, we discuss three related but physically distinct phenomena: inherent peculiar velocities of the neutral hydrogen atoms, and then the observational effects of redshift space distortions and the lightcone. Peculiar velocities are the motions of matter relative to the smooth Hubble expansion, arising from gravitational infall, outflows, and other dynamical processes. These velocities shift the observed frequency of the 21 cm line, producing the observational effect of redshift-space distortions (RSDs), apparent displacements of structure along the line of sight when distances are inferred from redshift. RSDs cause the Kaiser effect \citep{Kaiser_RSD_1987}, which describes the effect of dense regions appearing denser and underdense regions appearing more diffuse than they are due to hydrogen atoms moving towards overdensities and away from underdensities. The lightcone effect is a observational phenomenon due to the fact that different redshifts correspond to different cosmic times, and so the 21 cm signal evolves along the line of sight. \citep{Barkana_2006, Datta_lightcone_2012, Datta_2014, La_Plante_2014, Zawada_LC_2014, Ghara_lightcone_effect_2015}. The lightcone effect has been found to enhance the power at large scales and suppress it at small scales, with the crossover frequency evolving with redshift \citep{Datta_lightcone_2012}.

Semi-numerical models, such as SimFast21 \citep{Santos_2010} and 21cmFAST \citep{Mesinger_21cmfast_2011}, are often used to simulate the $21$ cm signal during the EoR and Cosmic Dawn. It is standard practice in such codes to create lightcones of $21$ cm brightness temperature by stitching together slices from $21$ cm brightness temperature boxes at set redshifts (known as coeval boxes), where the z axis of the lightcone corresponds to the observed frequency of the signal (this axis is known as the line-of-sight axis). Coeval boxes are 3D realisations of the $21$ cm brightness temperature with the same light travel time for all slices within the box. However, the brightness temperature is purely an observed quantity, so it is technically unphysical to consider coeval boxes of this quantity. 

Many existing approaches apply peculiar velocities to a coeval cube to generate redshift-space distortions, and only afterwards impose the lightcone effect by assembling slices from different redshifts or vice versa. This assumes that the two effects are independent. In reality, they are coupled: peculiar velocities, and thus the resulting RSDs, depend on the local redshift, and their mapping into observed frequency space changes along the line of sight. Previous works such as \cite{Mondal_lightcone_effect_2018} and \cite{Pramanick_resolution_lightcone_2023} have also attempted to incorporate light-cone construction and redshift-space distortions within the same framework, applying peculiar velocities to particles mapped onto the evolving light-cone. While conceptually similar, their implementation remains distinct from the fully coupled, self-consitent formulation developed in this study. Our method applies both effects simultaneously, ensuring that RSDs are computed with the correct velocity field for each redshift slice within the evolving light cone,  reflecting both the cosmic evolution and the velocity-induced frequency shifts.

We first presented this method in \cite{Chapman_Santos_2019}, finding a $60\%$ decrease ($80\%$ increase) in power over the largest (smallest) scales on the sky and a longer tail of bright temperatures in the brightness temperature distribution. In addition, \cite{Chan_rsd_24} found a difference of 5 \% to $<$ 10\% to arise in the global $21$ cm signal between the two methods, and \cite{wu_cavities_2024} found spectral features to be missed when using the standard approach.

In this paper, we build on work from \cite{Chapman_Santos_2019}, which we hereafter refer to as CS19, by further investigating the combined RSD and lightcone effects in a self-consistent manner. In addition, we assess the observability of the effects after the addition and removal of diffuse radio foregrounds and instrumental noise. This paper is organised as follows: Section~\ref{method} contains a brief outline of our approach to simulating a more robust lightcone and quantifies differences caused by each assumption made in the standard method. Section~\ref{resolution} contains an in-depth look at the redshift, frequency, and spatial resolution necessary for an accurate lightcone. Section~\ref{observability} shows the recovery of the $21$ cm signal from both methods via non-parametric foreground removal, after the addition of diffuse foregrounds and instrumental noise to the $21$ cm data. 

This paper has used a Planck $\Lambda$CDM cosmology with $\Omega_{\rm m}=0.315$, $\Omega_\Lambda=0.6847$, $\Omega_{\rm b}=0.0493$, $h=0.674$, $\sigma_8=0.811$ and $n=0.965$ \citep{Planck_results_2020}.

\section{Methods}
\label{method}

We will now give a brief outline of how our lightcone is simulated, taking into account the peculiar velocities of the hydrogen atoms. First, we will outline what we refer to as the basic method. These are methods which assume that the peculiar velocity and lightcone effects are decoupled; for more details, refer to CS19.

\begin{figure*}
    \begin{centering}
    \includegraphics[width=\textwidth,trim=0 0 0 5cm,clip]{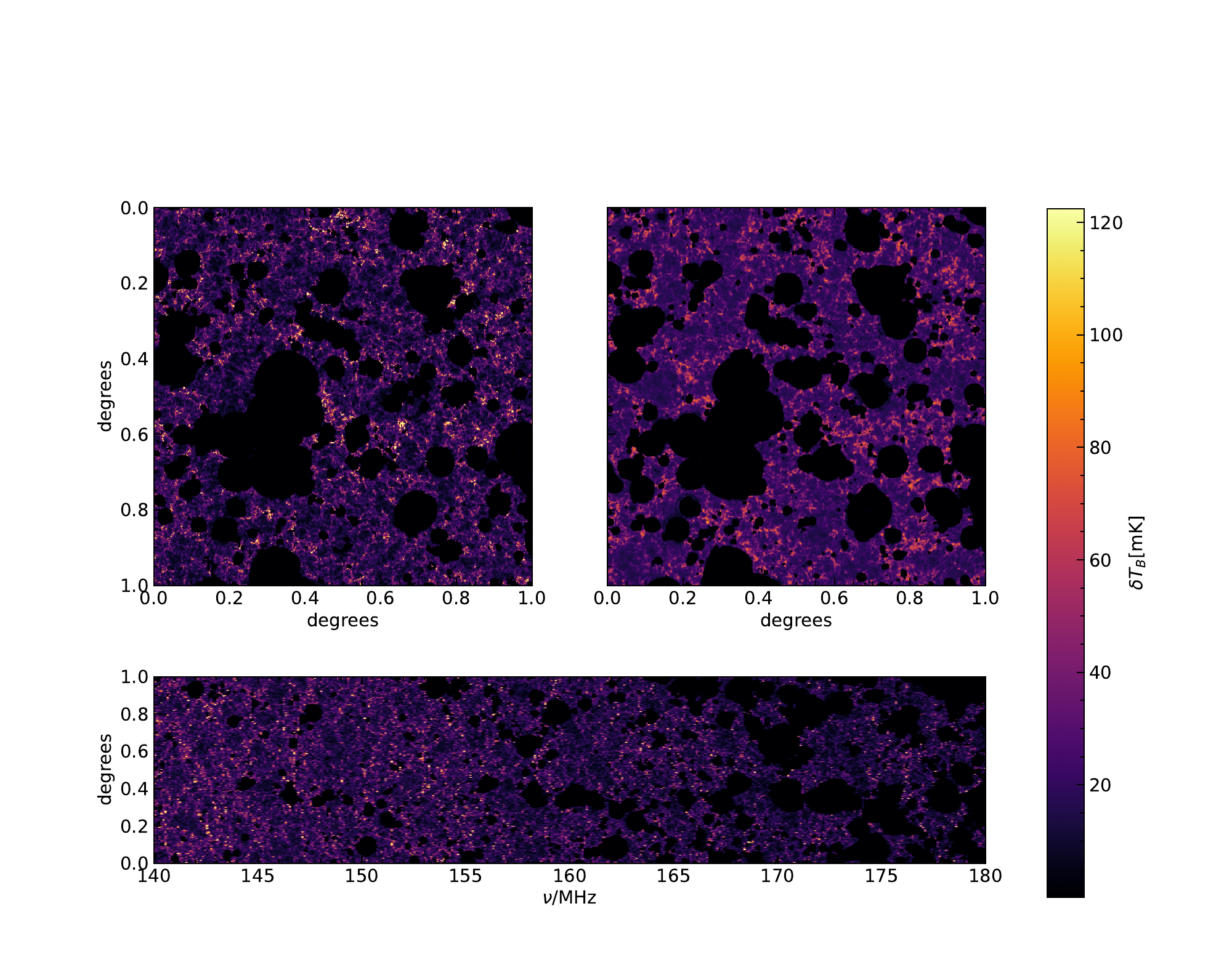}
    \vspace{-1cm}
    \caption{{\it{Top:}} Slices of the simulated 21cm brightness lightcone at 155 MHz. {\it{Left:}} Our extended approach to generating a lightcone and {\it{right:}} the basic  lightcone approach. {\it{Bottom:}} a slice of the extended lightcone generated between 140-180MHz. All images have been normalised to be on the same scale and saturated at the maximum value for the basic lightcone at 155MHz to better visualise the differences between the two methods. It is clear that the extended method results in brighter temperatures on small spatial scales.}
    \label{fig:visualing}
    \end{centering}
\end{figure*}
\begin{figure*}
\begin{centering}
    \includegraphics[width=0.48\textwidth]{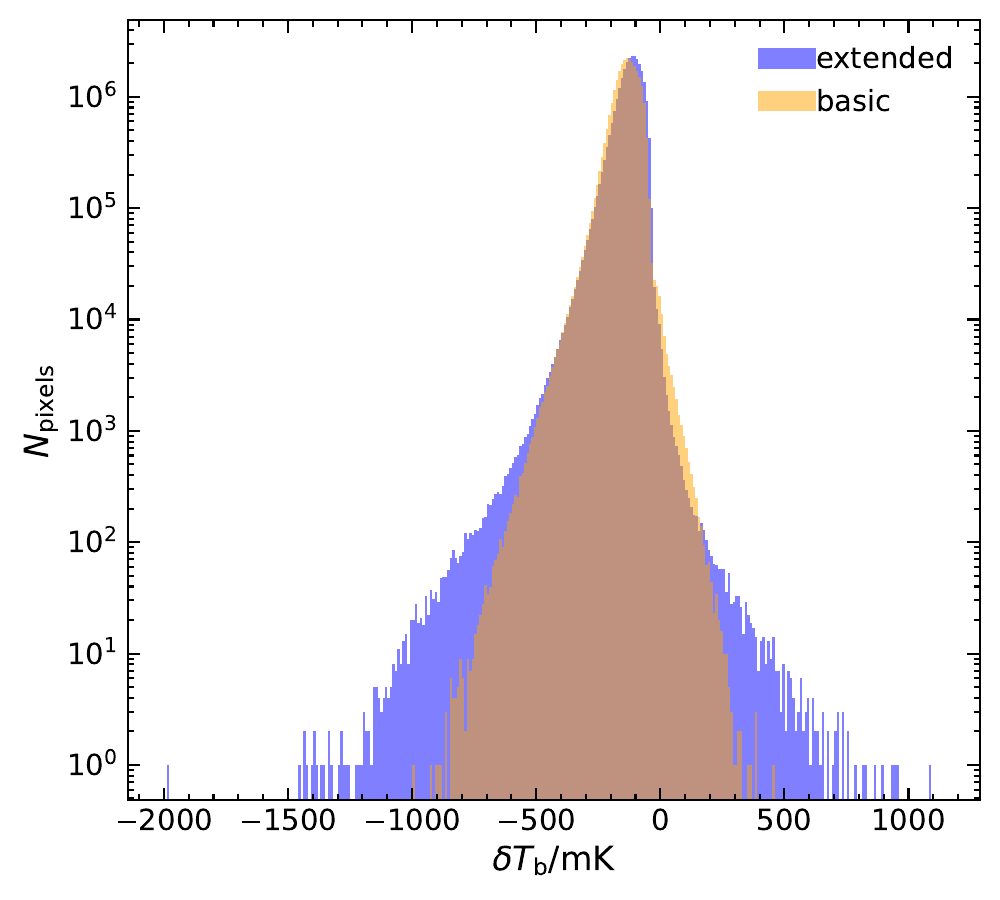}
       \includegraphics[width=0.48\textwidth]{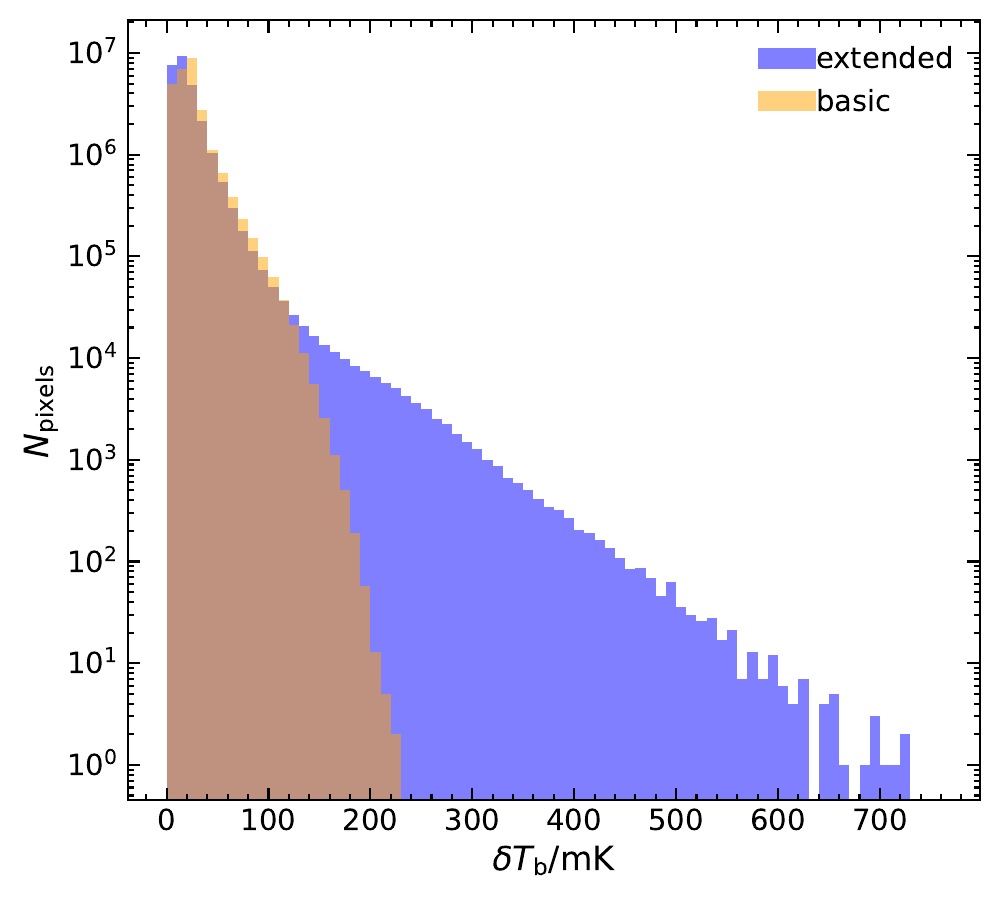}
    \caption{The voxel distribution of $21$ cm brightness temperatures ($\delta T_{\rm b}$) for our extended lightcone method (blue) and a basic lightcone method (yellow). Both distributions have been created from a 10 MHz lightcone between ({\it left}) 80-90 MHz ($x_{\rm HII} =0.000-0.001$) and ({\it right}) 150-160 MHz ($x_{\rm HII} =0.3-0.4$), using a bin spacing of 10 mK. In both frequency ranges, the distribution has an elongated tail of bright voxels when using our extended method.}
    \label{fig:Tb_hist}
    \end{centering}
\end{figure*}

\subsection{The basic method}
\label{sec:basic}
The $21$ cm signal at frequency, $\nu_0$, is measured by the brightness temperature, $\delta T_b(\nu_0)$, which is defined relative to a background intensity. It is standard to use the CMB temperature today ($T_{\rm CMB}(z=0)$, $2.725$ K) due to the rarity of strong background sources. We can ignore temperature fluctuations in the CMB due to the much larger fluctuations in brightness temperature, hence using a constant value of $T_{CMB} = 2.725$ K. The brightness temperature is defined as,
\begin{equation}
\label{eq:T21}
    \delta T_b(\nu_0) \equiv I_{21}(\nu_0)\frac{\lambda_0^2}{2k_B}-T_{\rm CMB}(z=0)
\end{equation}

Where $I_{21}(\nu_0)$ defines the intensity of $21$ cm radiation at $\nu_0$, $\lambda_0$ corresponds to the wavelength of the $21$ cm photon at that frequency, and $k_{\rm B}$ is the Boltzmann constant. Contributions to the $21$ cm brightness temperature can arise from three main mechanisms: (1) absorption/scattering of $21$ cm photons along a patch of neutral hydrogen, (2) the emission of $21$ cm radiation from a patch of neutral hydrogen, and (3) from peculiar velocities and Doppler effects. In which we refer to a 21cm contribution as an emission or absorption feature along an infinitesimal line element that contributes to $\delta T_b$.

In literature it is usual to make the following assumptions:

\begin{itemize}
    \item The $21$ cm line profile is taken to be a small-width top hat function, centered at the observed $21$ cm frequency, $\nu_{\rm obs}= \frac{\nu_{21}}{z + 1}$. Where $\nu_{21}=1420$ MHz is the rest frequency of the $21$ cm event.
    \item There is assumed to be only one contribution per observed frequency, per area of the sky. 
    \item $|dv/ds|/H << 1$ and  $\Delta \nu_{21}/\nu_{21} <<1$, as well as $j_{21}$ and $k_{21}$ being constant across the line width of the emission.
\end{itemize}
Therefore, the $21\rm cm$ brightness temperature equation (Eq.\ref{eq:T21}) can be approximated as,
\begin{equation}
\label{21cm_approx}
    \delta T_B (\nu_0) \approx (1.82 \times 10^{-28} \rm K m^2) \frac{n_{\rm HI}}{1 + z}\times\frac{cH^{-1}}{1 + 1/H \frac{dv}{ds}}\left[ 1 - \frac{T_{\rm CMB}}{T_{\rm S}}\right]
\end{equation}
Where $n_{\rm HI}$ is the number density of $\rm HI$ and $T_{\rm S}=\frac{j_{21}}{k_{21}}\lambda_{21}^2/(2k_{\rm B})$ is the spin temperature of $21$ cm photons. As can be seen from Eq.~\ref{21cm_approx}, an infinity can occur when $\rm \frac{dv}{ds} \rightarrow - H$, so it is standard to set a threshold to prevent $\rm \frac{dv}{ds}$ approaching this limit.

Coeval brightness temperature boxes are first simulated using Eq.\ref{21cm_approx}. The boxes are equally spaced in redshift over the range required for the desired lightcone. For an observed frequency $\nu_i$, the corresponding 21-cm signal arises from a redshift
\begin{equation}
1 + z_i = \frac{\nu_{21}}{\nu_i}
\end{equation}
and a comoving distance $x_i$.  
We load the coeval cube at redshift $z_i$ and extract the slice at comoving distance $x_i$. The corresponding cell index is  
\begin{equation}
k_i = \frac{x_i - x_{\mathrm{min}}}{\Delta r},
\end{equation}
where $x_{min}$ is the comoving distance to the near edge of the lowest redshift coeval cube, and $\Delta r$ is the comoving cell spacing. This slice, located at the line-of-sight index $k_i$, is the only contribution to the observed-frequency map at $\nu_i$.

Periodic boundary conditions along the line-of-sight can be used to ensure a pixel is always found, even in the event that $x_i$ becomes much larger than the coeval box size. Interpolation between coeval boxes can also be used to more accurately match the lightcone redshift-frequency equivalency. \cite{Pramanick_resolution_lightcone_2023} showed that using interpolation can still give an accurate lightcone provided adequate redshift spacing is used between coeval boxes.

There are methods that include peculiar velocity effects as a post-processing step on coeval brightness temperature boxes \citep[e.g.][]{Mao_RSD_2012,Jensen_RSD_2013,Datta_2014}, for example using cell-shuffling. These methods often make one or more assumptions, such as not allowing multiple 21 cm events along the line of sight, imposing a threshold on $1 + 1/H \frac{dv}{ds}$, or saturating the spin temperature, $T_s>>T_{CMB}$. For a discussion of these methods, see CS19 Section 4.2. We choose to compare our method to the most basic method using equations 3 and 4, in line with CS19, as our focus is whether the differences found in that paper can be recovered after foreground removal.

\subsection{The extended method}

Many existing approaches first apply peculiar velocities to a coeval cube to generate redshift-space distortions, and only afterwards impose the light-cone effect by assembling slices from different redshifts. This assumes that the two effects are independent. In reality, they are coupled: peculiar velocities — and thus the resulting distortions — depend on the local redshift, and their mapping into observed frequency space changes along the line of sight. Our method applies both effects simultaneously, ensuring that velocity-induced distortions are computed with the correct velocity field for each redshift slice within the evolving light cone.

Our approach differs as we fully model the peculiar velocities and redshift space distortions simultaneously, by taking into account line broadening and radiative transfer effects. We use Eq.~\ref{eq:T21} rather than Eq.~\ref{21cm_approx} to calculate the brightness temperature of each voxel (3D pixel) of our lightcone. By doing so, we allow for more than one $21\rm cm$ contribution to occur per voxel and remove the assumption that the background temperature is always $T_{\rm CMB}$. This also allows for a wider range of peculiar velocities $\left(\rm \frac{dv}{ds}\right)$ since we no longer have an infinity when $1/H \frac{d\nu}{ds} \rightarrow -1$, which can occur in high-density regions of space. 

When filling a voxel of $\nu_0$, we start at a high enough $z$ to ensure that there is no $21$ cm contribution before it. Considering each voxel of the lightcone individually, we integrate down the line-of-sight to the observer. We take small steps in proper distance to ensure an accurate lightcone is made. We calculate the frequency for each step from the previous step's frequency until we intersect with the frequency of the $21$ cm line, and we can define that there is a contribution to $\delta T_b$. This allows for the inclusion of partial intersections with the $21$ cm frequency. If no contribution is found then the intensity of that pixel is taken as the intensity of the CMB scaled by expansion. The final intensity is calculated as $I_{21}(\nu_0)=(\nu_0/\nu_{21})^3I_f(\nu_{21})$, where $I_f$ may be the result of multiple contributions to the $21$ cm brightness temperature.

Our method differs from what we define as the basic approach as we calculate the brightness temperature of each voxel along the line-of-sight from coeval boxes of the halo, ionisation, Lyman-$\alpha$, velocity, star formation rate, density, and X-ray fields, we will refer to these as `ingredient' boxes, rather than coeval boxes of already calculated $21$ cm brightness temperature, as this is not a physical quantity and only defined in observational space. For a more in-depth discussion of this method see CS19.

Fig.~\ref{fig:visualing} shows slices of the lightcones generated for the extended (left) and basic (right) methods in the top panel, alongside a slice along frequency of a lightcone made using the extended method in the bottom panel. This figure highlights the differences seen in the small-scale structures and that the extended method contains a higher number of brighter pixels. We see this to be the case regardless of frequency. To highlight this further, we show the histograms of the $21$ cm brightness temperature distribution for the basic (red) and extended (blue) lightcone methods in Fig.~\ref{fig:Tb_hist}. It can be seen that our extended method elongates the tail distribution, although note the logarithmic $y$-axis scale. The distribution is shown for two 10 MHz lightcones, between $80-90$ MHz (left) and $150-160$ MHz (right). In the $150-160$ MHz range, the maximum $\delta T_{\rm b}$ for the basic method is $225 $ mK and for our extended method is $769 $ mK. The percentage of the extended methods' distribution that sits $>225 $ mK is only $0.1\%$, but these correspond to the brightest pixels in the lightcone, and as we will see, this small percentage of voxels makes a large difference to the observed power spectra. The $80-90 $ MHz range shows a similar behavior extending the distribution to both positive and negative $\delta T_{\rm b}$, with the maxima and minima of the distribution sitting at $-998 <\delta T_{\rm b}/\rm mK < 460$ for the basic method and $-1989<\delta T_{\rm b} /\rm mK<1132 $ for our extended method. These extensions of the distributions account for 0.004\% of the voxels, but again, they are the brightest pixels and create large differences in the power spectra. More comparison graphs that highlight the differences between the two methods can be found in CS19.

\subsection{Quantifying the difference}

There are multiple effects that contribute to the differences seen between each method, and we now examine each of these in turn. 

\begin{figure}
\centering
    \includegraphics[width=\linewidth]{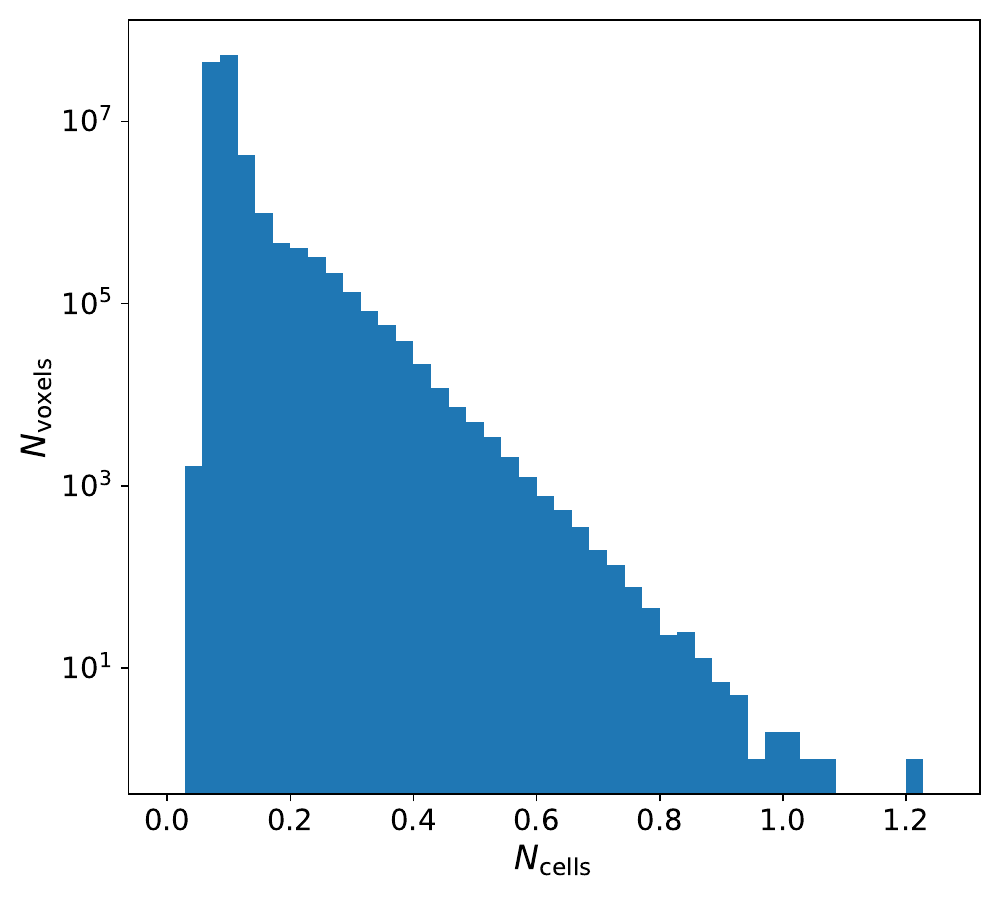}
    \vspace{-0.6cm}
    \caption{A histogram showing the number of steps which result in a contribution of 21 cm intensity, normalised by the number of steps the algorithm takes across a cell. The basic method would show all voxels to be comprised of $1.0 N_{\rm cell}$. Values above 1 indicate that there has been intensity added in addition to what the basic method would find. Values below 1 indicate the method finds only a partial contribution across the cell. This figure shows the distribution for a 10MHz lightcone between $150$ and $160$ MHz made using our extended method. With the extended method, we see an average of 8\% and a maximum of 120 \% of a cell per voxel.}
    \label{fig:one_event}
\end{figure}  

\begin{figure*}
    \includegraphics[width=\textwidth]{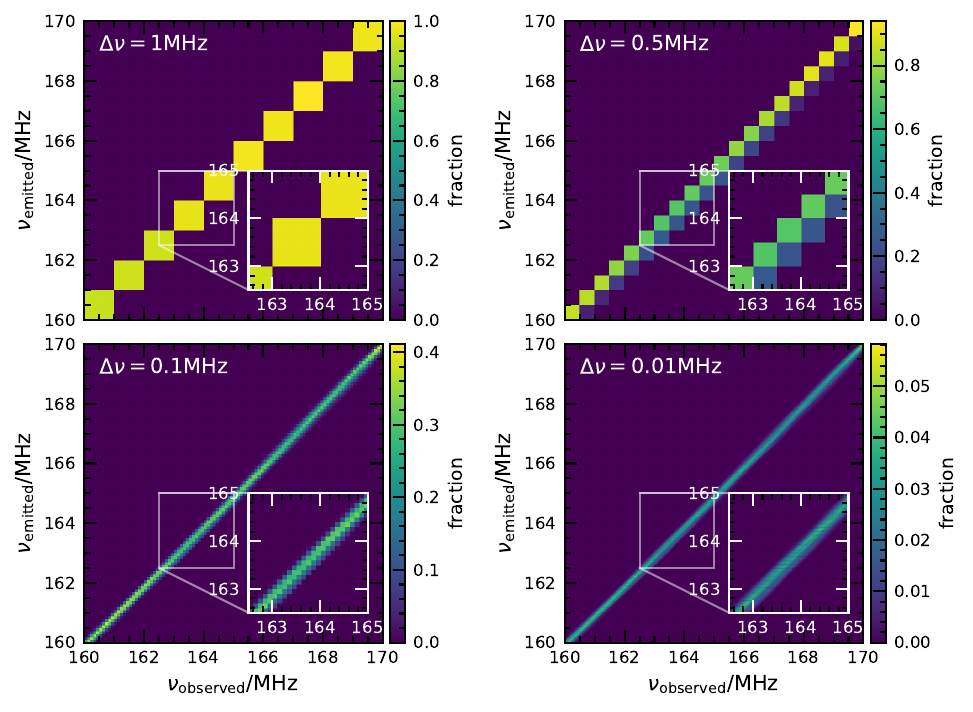}
    \caption{All panels show the frequency that the observed $21$ cm photon, $\nu_{\rm observed}$, appears in the lightcone versus the frequency slice of the coeval box the photon was emitted from, $\nu_{\rm emitted}$. The different panels show different frequency resolutions of lightcones (clockwise from top left) $\Delta \nu = 1, 0.5, 0.1, 0.01$. In all plots, the graphs are colour coded by the number of contributions, and normalised by the number of voxels per slice. Each plot shows a zoomed-in region of the graph. We see a smoothing out of the distribution when approaching finer frequency spacing.}
    \label{fig:slice-counter}
\end{figure*}
\begin{figure*}
    \includegraphics[width=0.99\textwidth]{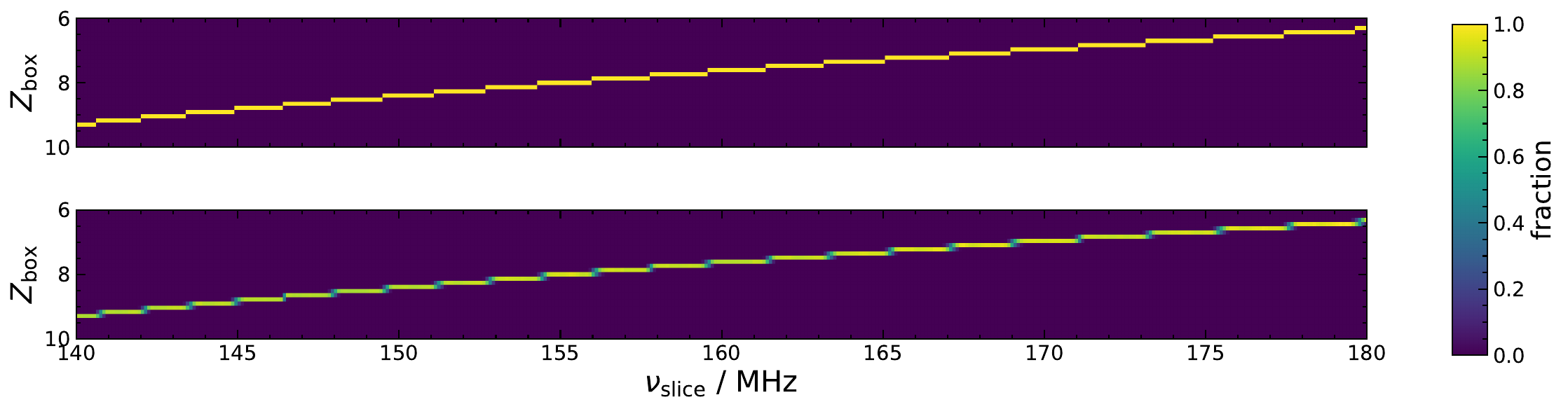}
    \caption{Both panels show coeval box redshift from which a $21$ cm contribution arises plotted against the frequency of the lightcone slice that the contribution is observed at. We have coloured coded this plot by the number of contributions per frequency, normalised by the number of voxels per slice. The basic method is shown in the {\it{top}} panel, and our extended method is shown in the {\it{bottom}} panel. We see an overlap of where more than one coeval box contributes to a lightcone slice at frequencies within $\pm 0.25$ MHz of where the basic method changes redshift box. The same colourscale is used for both panels to allow an easy comparison to be made.}
    \label{fig:zbox_nu_rsd}
\end{figure*}
\subsubsection{Partial Contributions}
The first is that due to the fact that our step size, $ds$, is less than the coeval box cell length, we allow for partial emission or absorption features. Additionally, we do not impose a limit on the number of steps that contribute to a voxel of the lightcone, thus potentially allowing contributions of intensity from more than one coeval cell. The basic method we defined in Section.~\ref{sec:basic}, and most methods that incorporate peculiar velocity effects, assume that $I_{\rm 21}$ in Eq.~\ref{eq:T21} occurs due to only one cell of the $21$ cm brightness temperature coeval box per voxel of the lightcone. However, multiple contributions to a lightcone voxel are physically possible due to photons red or blue-shifting into the observed frequency range by peculiar velocities or Doppler effects due to the expansion of the Universe.

In Fig.~\ref{fig:one_event}, we show a histogram of the fraction of a coeval cell that contributes to each voxel in a 10MHz lightcone, made using our extended method. We can think of this as the number of steps that result in a contribution of 21 cm intensity, normalised by the number of steps the algorithm takes across a cell. Values above 1 indicate that there has been intensity added in addition to what the basic method would find. Values below 1 indicate the method find only a partial contribution across the cell, compared to the basic method which would again assume the whole cell contributes. From this, we can see that on average each voxel of the lightcone consists of 8\% of a coeval cell, and only five voxels consist of 100\% or more of a cell. For comparison, the basic method's distribution would show all voxels to be comprised of 100\% of a coeval cell.

The voxels with contribution fractions of above 1.0 were seen to 
have the largest differences in brightness to corresponding voxels in the basic lightcone.

\subsubsection{Frequency Shifting}
We looked at the red/blueshifting that can occur from photons emitted at one frequency being observed at a different frequency. We wanted to investigate the frequency range in which this shifting can occur when using our extended method. The basic method matches an observed lightcone frequency with a single brightness temperature slice, either directly from a coeval box or interpolated between boxes. Methods that include peculiar velocity effects make the same assumption that one slice of a coeval box is used per frequency slice of the lightcone. Fig.~\ref{fig:slice-counter} shows this effect for four different frequency resolutions of our extended method's lightcone. Clockwise from left we show $\Delta \nu = \,1.0, \,0.5,\, 0.1,\,0.01 $ MHz. The colourbar is normalised as such that it shows the fraction of light in the observed lightcone slice that comes from each $\nu_{\rm emitted}$. We note that the basic method would show the same trend as the extended method's $\Delta \nu = 1.0$ for every frequency resolution. We can see that $21$ cm photons are being shifted to different frequencies within a $ 1$ MHz range; therefore, at frequency resolutions $\geq 1 $ MHz we would not see any shifting. We also see that the maximum value of the colourscale decreases with increasing frequency resolution. This is because there are more voxels contributing to a $1$ MHz pixel range, so the number of voxels per coeval slice is less. We also see that there is a frequency dependence to the amount of shifting that occurs, showing that the combined effect of peculiar velocities and the lightcone effect is frequency dependent.

We also looked at allowing multiple coeval boxes to contribute to a $21$ cm event. For the basic method's lightcone, the coeval brightness temperature box closest to $z = \frac{\nu_{21}}{\nu_{i}}-1$ will contribute to the lightcone slice, and the slice will be interpolated if it is not exactly $\nu_i$. For our extended method, as we move along the line-of-sight, we load coeval boxes at different redshifts to assess for $21$ cm contributions. While most contributions will arise from the equivalent wavelength to the observed frequency, there may be some contributions that arise from other redshift boxes. This is because the gradient of the peculiar velocity shifts the observed frequency as in Eq.~\ref{eq:T21}. This effect is visualised in Fig.~\ref{fig:zbox_nu_rsd}, where we plot the redshift of the coeval box used against the observed frequency slice of the lightcone. The top panel shows the basic method, and the bottom shows our extended method. The colourbar shows the fraction of the lightcone slice which arises from the corresponding redshift coeval box. From this, it can be seen that multiple redshift coeval boxes contribute to a slice of the lightcone when the observed frequency is within $\pm 0.2$ MHz of where the basic  method changes coeval boxes. This is because $\frac{dv}{ds}$ is relatively small. We see that this effect would increase with increased coeval box resolution, and introduces a `fuzziness' when considering which coeval voxels contribute to a lightcone slice, in comparison to the single slices taken from coeval boxes when using the basic method.

\section{What Resolution is needed for an accurate lightcone?}
\label{resolution}

\begin{figure}
\begin{centering}
    \includegraphics[width=0.5\textwidth]{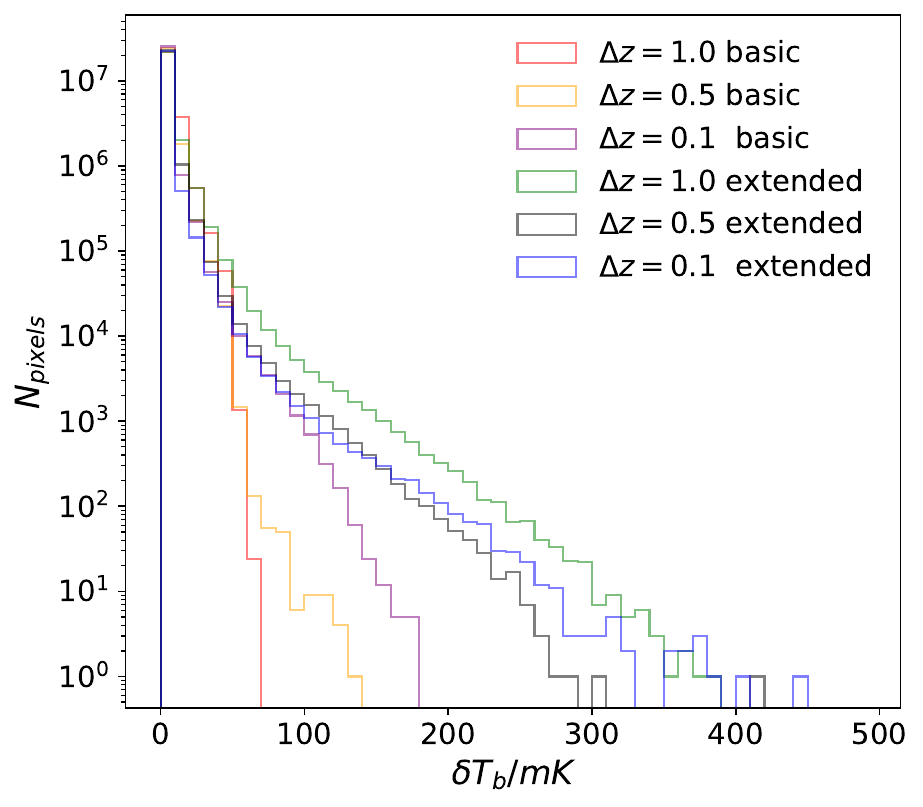}
    \vspace{-0.4cm}
    \caption{A histogram of the voxel $21$ cm brightness temperature distribution ($\delta T_{\rm b}$) for a $10$ MHz lightcone centered at $185$ MHz ($x_{\rm HII}=0.85$). Three different coeval box spacing resolutions are shown $\Delta z = 1.0, 0.5,0.1$ in red, yellow, and purple for the basic method and green, black, and blue for our extended method. We do not see a convergence reached at these coeval box spacing resolutions.}
    \label{fig:zbox_pixel_hist}
    \end{centering}
\end{figure}
\begin{figure*}
\begin{centering}
    \includegraphics[width=0.99\textwidth]{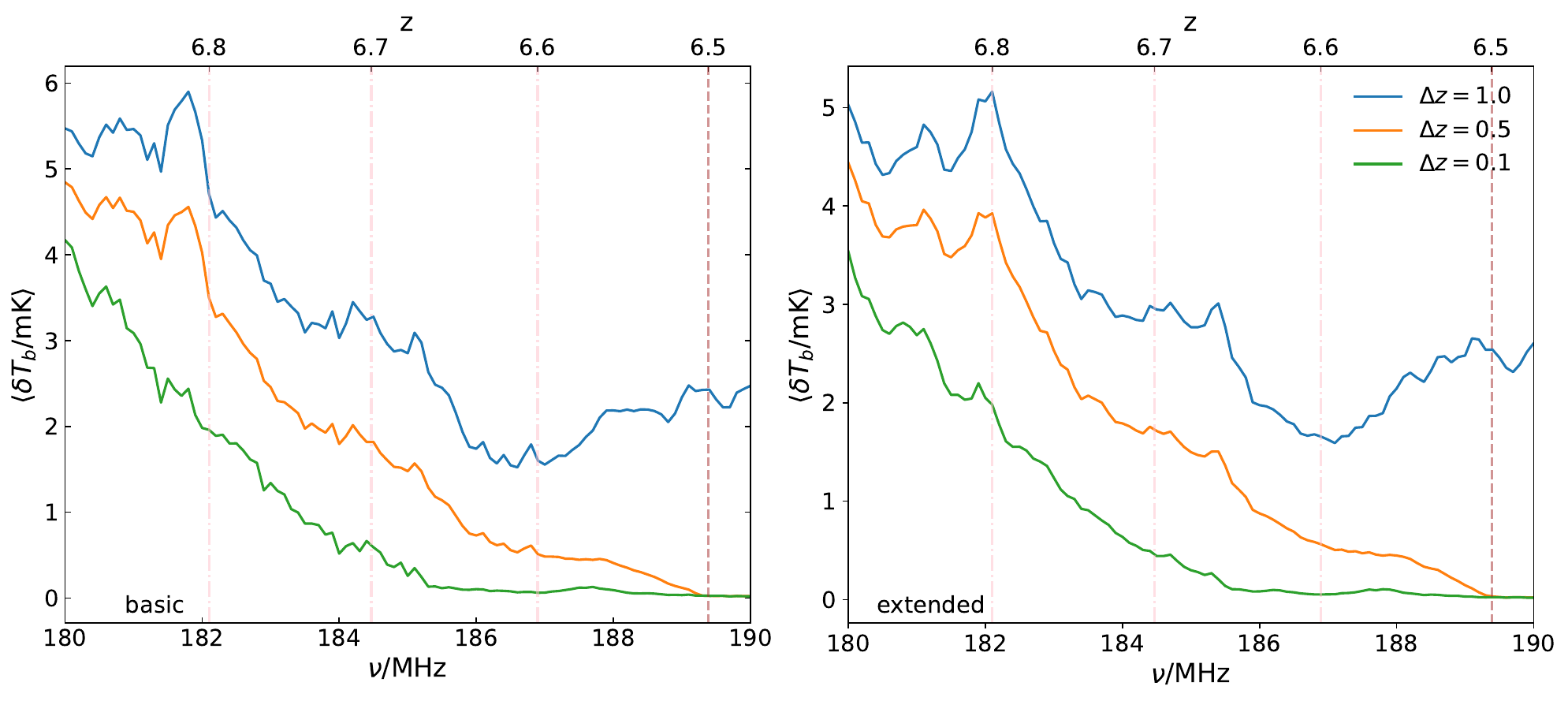}
    \vspace{-0.4cm}
    \caption{The globally averaged $21$ cm brightness temperature, $\langle \delta T_{\rm b} \rangle$, as a function of redshift. This is shown for $\Delta z = 1.0,0.5,0.1$ in blue, orange, and green. The basic method ({\it{left}}) and our extended method ({\it{right}}) are both shown alongside vertical pink dot-dashed lines indicating every $0.1 \Delta z$ and brown dashed lines highlighting every $0.5 \Delta z$. We see a convergence of the endpoint of reionisation towards smaller redshift spacing, but note that finer spacing than we have used here would be needed for complete convergence, although this would be computationally costly.}
    \label{fig:zbox_global}
    \end{centering}
\end{figure*}
\begin{figure*}
\begin{centering}
    \includegraphics[width=0.999999\textwidth]{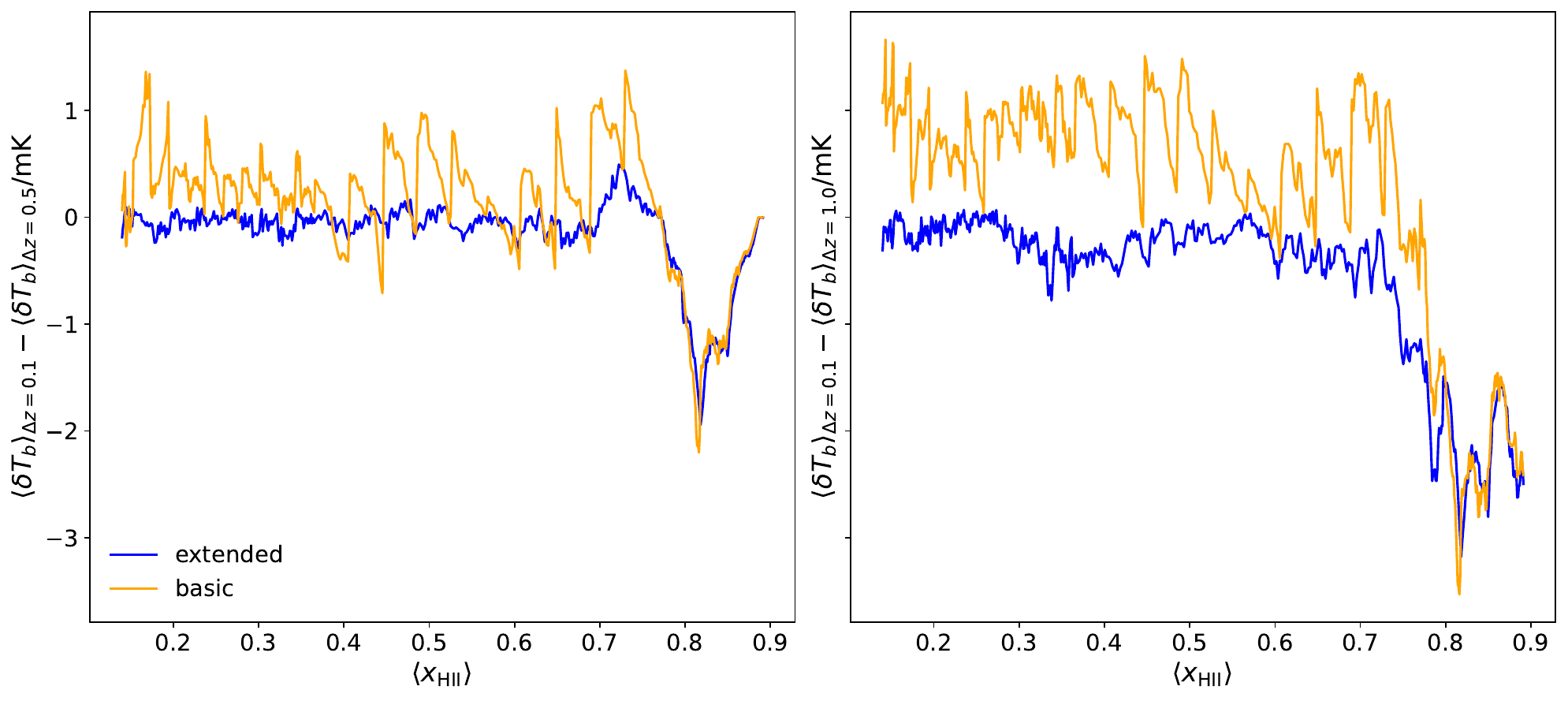}
    \vspace{-0.4cm}
    \caption{The difference in the globally averaged $21$ cm brightness temperature, $\langle \delta T_{\rm b} \rangle$, between  $\Delta z =0.5$ ({\it{left}}) and $\Delta z =1.0$ ({\it{right}}) and our fiducial spacing ($\Delta z =0.1$) as a function of ionised hydrogen fraction ($\langle x_{\rm HII} \rangle$). We note that for both spacings the extended method is more converged and that at higher ionisation fractions the convergence worsens.}
    \label{fig:zbox_global_evo}
    \end{centering}
\end{figure*}
When creating a lightcone, regardless of the method choice, the chosen resolution can have a big effect on the output lightcone and computational costs. We turn to look at the effect of redshift, frequency, and spatial resolution, as well as the limit at which the basic  and extended lightcone methods converge. This is important to consider, as the differences between the two methods will be resolution-dependent; therefore, we want to pick a resolution that allows us to fully investigate these effects. The importance of resolution choice was shown in \cite{Datta_lightcone_2012}, who found the lightcone effect to scale with redshift bin size.

\subsection{Redshift resolution}

Firstly, we look at the effect of redshift spacing between coeval boxes on the lightcones produced. Making these coeval boxes can be computationally costly and require large storage, so it is important to optimise the number made. In our extended method, redshift snapshots of all the `ingredient' boxes are made and these are used to calculate the $21$ cm signal in the lightcone along each line-of-sight, as in Eq.~\ref{eq:T21}. In the basic  method, coeval boxes of the $21$ cm brightness temperature are also created, and slices of these boxes are used to create the lightcone. When using these coeval boxes, we can interpolate between redshift snapshots to reduce the number of boxes needed to reach a convergence and cut down on computational costs. Similar resolution effects have been examined by \cite{Pramanick_resolution_lightcone_2023}, who investigated how the separation between coeval snapshots influences the accuracy of light-cone realisations and proposed an interpolation scheme to mitigate this. In our work, this dependence is revisited primarily to quantify its impact on foreground mitigation and parameter recovery rather than to optimise the shapshot spacing itsef.  

In Fig.~\ref{fig:zbox_pixel_hist} we show the histogram of $21$ cm brightness temperature for a $10$ MHz lightcone centred at $185$ MHz ($x_{\rm HII}=0.85$), for three different coeval box spacings for both methods. All of the lightcones used for Fig.~\ref{fig:zbox_pixel_hist},~\ref{fig:zbox_global} and ~\ref{fig:zbox_global_evo} have $\frac{\theta}{N_{\rm pixel}}= 7 ''$ and $\Delta \nu = 0.1$. For all the coeval box spacings we have investigated, $\Delta z= 1.0, 0.5,0.1$, we recover the differences between the two methods. However, we see a noticeable distinction between the output for different coeval box spacings, which spans $\approx100 $ mK regardless of the method used. This is further highlighted by looking at the evolution of the global temperature in Fig.~\ref{fig:zbox_global}. In this figure, the global evolution of the $21$ cm brightness temperature is shown for the three redshift box spacings investigated, with the redshifts at which the boxes change highlighted by horizontal lines. From this, we see that regardless of the method used, there is a disparity between each resolution choice, and that the choice made will affect the endpoint of reionisation that each lightcone predicts. We note that although a convergence was not reached at $\Delta z = 0.1$, it was too computationally costly for us to simulate a smaller coeval box spacing. 

Fig.~\ref{fig:zbox_global_evo} shows the difference between the global temperature from our fiducial lightcone ($\Delta z = 0.1$) and a comparison lightcone. $\Delta z = 0.5$ is shown on the left and $\Delta z = 1.0$ on the right. These residual graphs are shown for both the basic (orange) and extended (blue) methods. The repetitive pattern seen in the residuals of the basic method is an artifact of taking the slices directly from each coeval $21$ cm brightness temperature box. When the same coeval box is chosen for the fiducial and comparison lightcones and there is no interpolation, the same slice of the $21$ cm brightness temperature coeval box is used to build up both lightcones. For all coeval box spacings, we see a stronger agreement with low ionisation fractions and when using our extended method. However, for both methods and resolution choices, we see a drop-off towards the end of reionisation. This drop off occurs at a lower ionisation fraction for sparser resolutions. For $\Delta z = 0.5 $ we see an increase in comparability of the two lightcones after $x_{\rm HII} = 0.82$, this is due to both spacings predicting a similar end point to reionisation. Whereas $\Delta z = 1.0 $ predicts a vastly different endpoint, so we do not see the same turnaround.

\subsection{Spatial resolution}
\label{sec:spatial_resolution}

\begin{figure}
\centering
    \includegraphics[width=\linewidth]{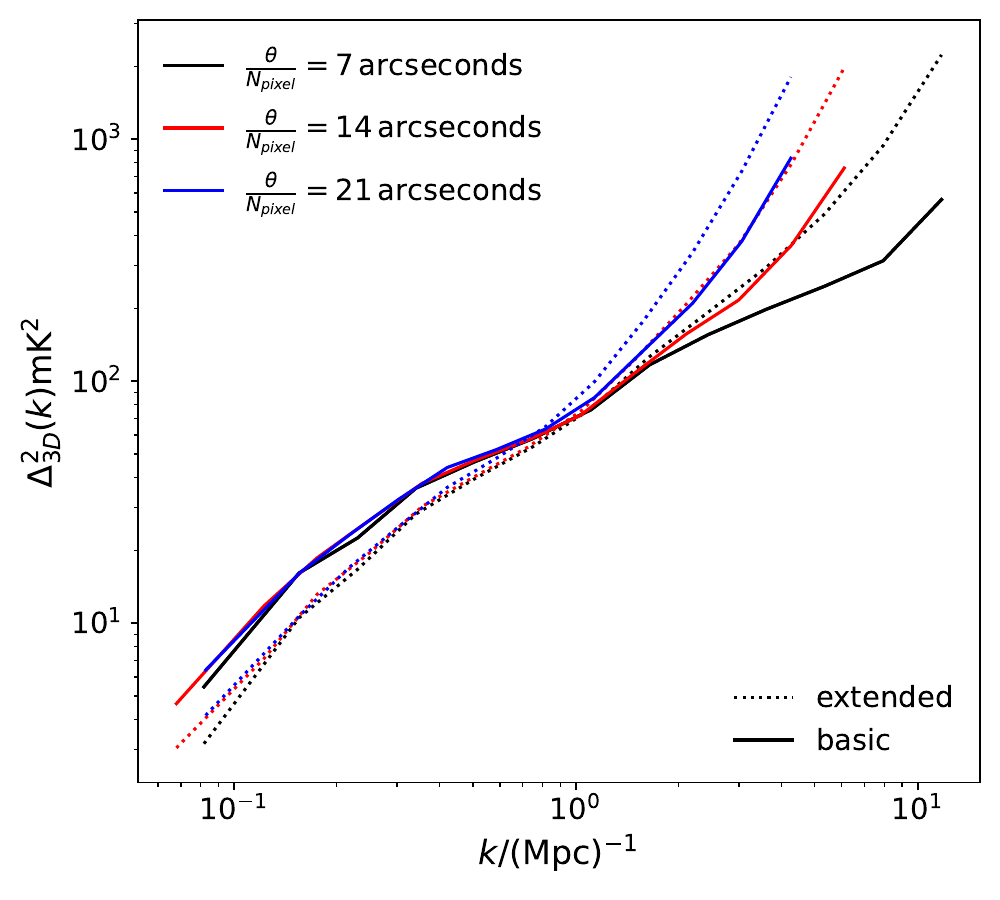}
   \vspace{-0.4cm}
    \caption{The 3D averaged power spectrum decomposed in $k$, $\Delta^2_{\rm 3D}(k)$, for three different spatial resolutions of lightcone, $\frac{\theta}{N_{pixel}}= 7,\,14,\,21\,''$ in black, red and blue respectively. Our extended lightcone method is shown in the dotted lines, and the basic method is shown in the solid lines. These power spectra have been taken from $10$ MHz lightcones centered at $145$ MHz. We see the enhancement in large $k$ to exist for all spatial resolutions, and to become more prominent as the resolution increases.}
    \label{fig:spatial_res}
\end{figure}

We now turn to look at the effect of different spatial resolutions on the power spectra for both lightcone methods. The 3D averaged power spectrum decomposed in $k$ is shown in Fig.~\ref{fig:spatial_res} for the three different spacial resolutions tested ($\frac{\theta}{N_{pixel}}=7,\, 14$ and $21'' $), all the lightcones used for this figure have $\Delta z=0.1$ and $\Delta \nu = 0.1$. These resolutions were chosen as $\frac{\theta}{N_{pixel}}=7 ''$ is our resolution limit due to coeval box resolution limitations. The extended method (dotted lines) and the basic method (filled lines) are shown in this figure for each resolution tested. It can be seen that the power spectra start and end at different $k$-scales for each resolution. This is because we varied the box size but kept the number of voxels the same to vary the resolution. For example, the largest box size corresponds to the coarsest resolution; therefore, smaller $k$ bins are reached, but the largest $k$ bins are lost. From this figure, we can see that the distinction between the two methods persists for all $k$ regardless of spatial resolution, especially the enhancement seen at large $k$ scales. This enhancement increases in magnitude as the resolution increases. This is due to our ability to more accurately resolve structure on the smallest available scales. We also see a slight resolution dependence on the turnover frequency and little to no dependence on the suppression at small $k$.

\subsection{Frequency resolution}

\begin{figure}
\begin{centering}
    \includegraphics[width=0.48\textwidth]{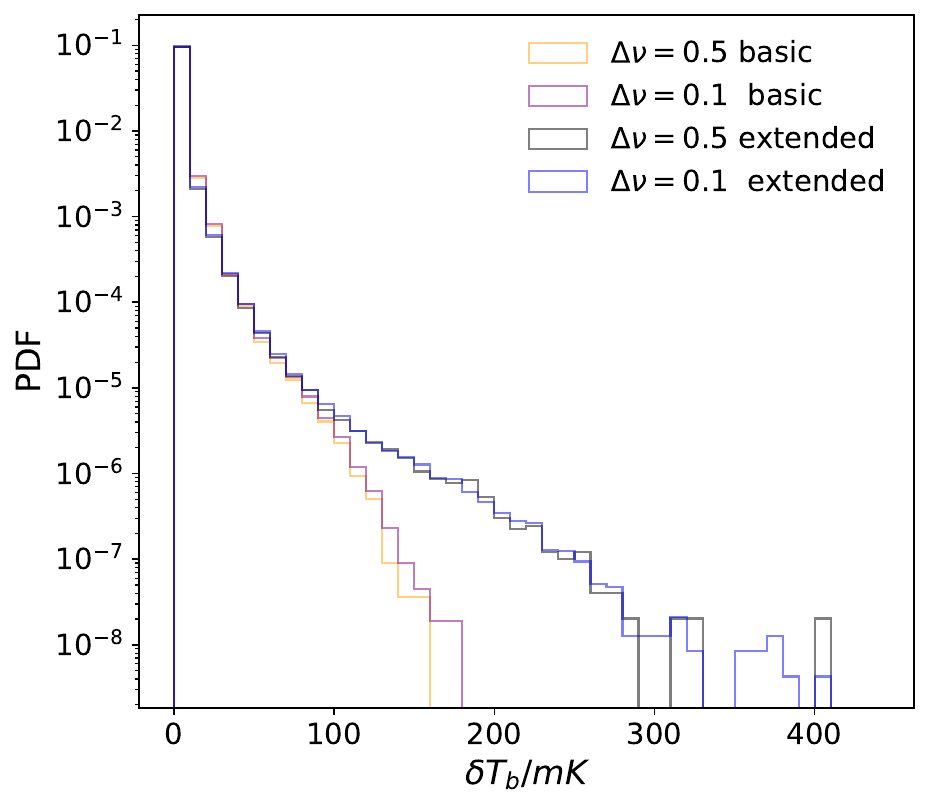}
    \vspace{-0.4cm}
    \caption{Probability density functions of the brightness temperatures of $21$ cm photons, $\delta T_{\rm b}$ from a $10$ MHz lightcone centered at $185$ MHz ($x_{\rm HII} = 0.85$). This is shown for lightcone frequency resolutions of $\Delta \nu = 0.5, 0.1$ for our extended method (black and blue) and the basic  method (yellow and purple). From this, we see that a convergence of frequency resolution has been reached.}
    \label{fig:nu_res_pdf}
    \end{centering}
\end{figure}
We now turn to investigate two different frequency resolutions, $\Delta \nu = 0.5$ and $0.1$, for both lightcone methods using $\Delta z = 0.1$ and $\frac{\theta}{N_{\rm pixel}}=7''$. We are focusing on these two frequency resolutions, as in Fig.~\ref{fig:slice-counter} we saw that a frequency spacing smaller than $1$ MHz was required to allow for peculiar velocity effects to red/blue-shift photons between frequency slices. Probability density functions (PDF) created from a $10$ MHz lightcone centred at $185$ MHz of different frequency resolutions are shown in Fig.~\ref{fig:nu_res_pdf}, where the PDFs have been normalised by the number of $21$ cm contributions in each lightcone. This figure shows that the shape of the distribution for each method is converged in frequency resolution. Hence, informing us that the redshift spacing of coeval boxes is more important to constrain reionisation scenarios than frequency spacing of the lightcones.

\section{Can a robust simulation of the peculiar velocity effect create an Observable difference?}
\label{observability}
\begin{figure}
\centering
    \includegraphics[width=\linewidth]{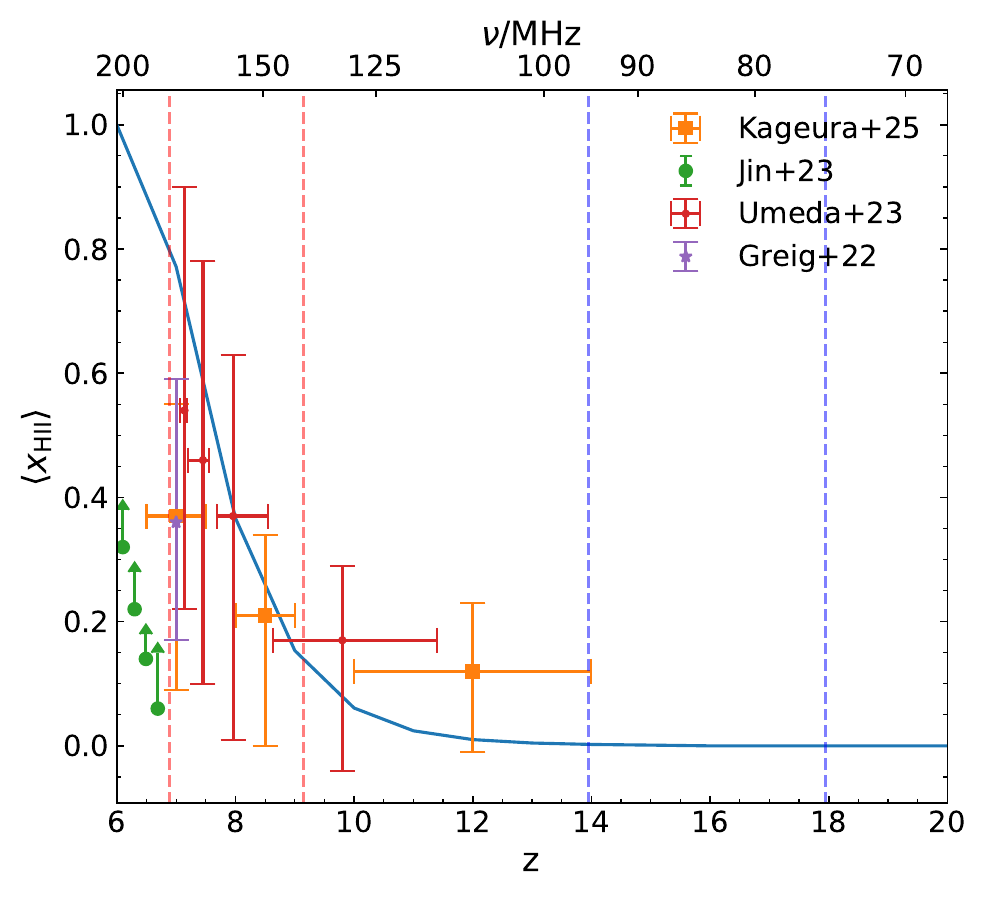}

    \caption{A plot of average ionised hydrogen fraction, $\langle x_{\rm HII} \rangle$, against redshift and the corresponding $21$ cm observed frequency. The reionisation history used in this paper is shown in blue alongside current observational constraints from Ly$\alpha$ equivalent widths \citep{Kageura_xHI_2025}, quasar damping wings \citep{greig_igm_2022}, galaxy damping wings \citep{Umeda_nhi_2023}, and Ly$\alpha$ dark pixel fraction \citep{jin_nearly_2023}. Dashed vertical lines have been drawn to show the frequency ranges of the lightcones used for our foreground removal analysis. The red dashed lines show the EoR lightcone ($140-180$ MHz) and the blue dashed lines show the Cosmic Dawn lightcone ($75-95$ MHz).}
    \label{fig:xHI}
\end{figure}

\begin{figure*}
\begin{centering}
    \includegraphics[width=0.48\textwidth]{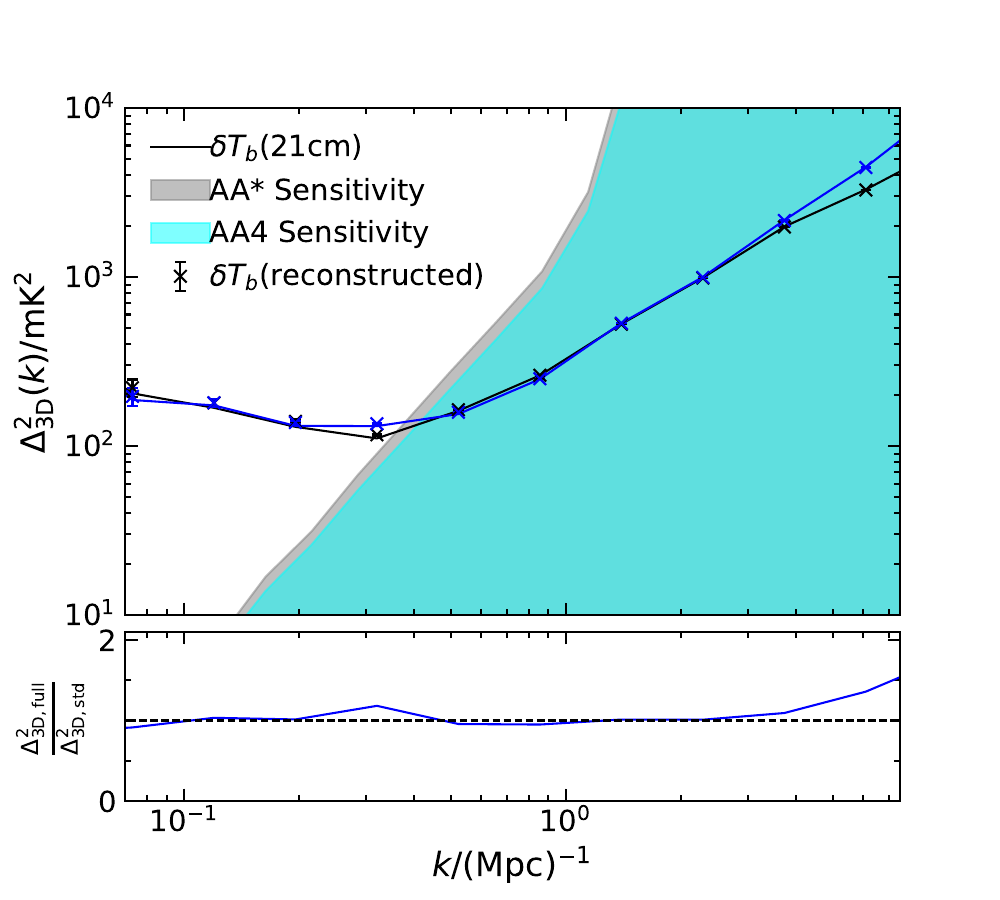}
    \includegraphics[width=0.48\textwidth]{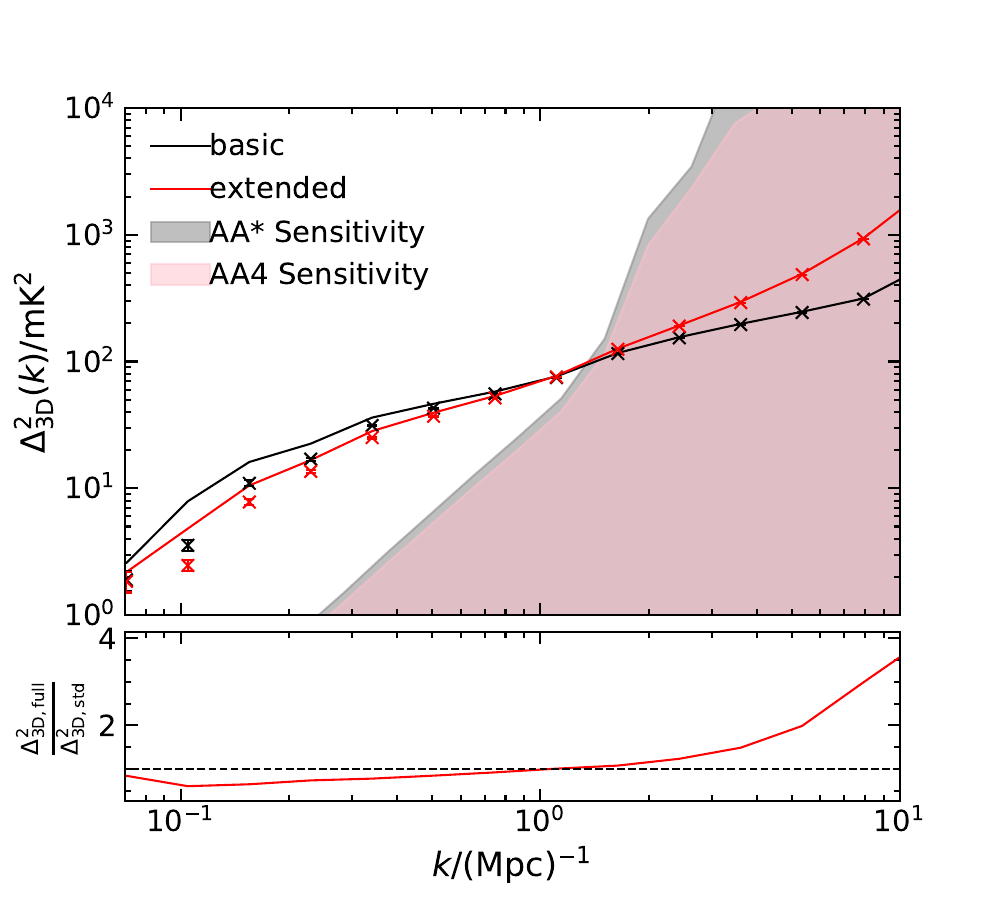}
    \vspace{-0.4cm}
    \caption{The 3D power spectrum decomposed into  $k$ for two 10$\rm MHz$ lightcones at {\it{left:}} 80-90 MHz and {\it{right:}} 150-160 MHz. Both panels show the true $21$ cm signal (solid lines) and the reconstructed signal (crosses) using a 4-component FastICA fitting. The coloured lines show when the extended approach has been used and the black lines when the basic  approach has been used. The grey (coloured) shaded area shows the sensitivity of a 1000-hour SKAO-low observation using the AA* (AA4) layout. The $k$ scale has been cut to reflect the sensitivity when using the full array with a maximum baseline length of 8km. The bottom panel shows the ratio of the extended to the basic method.}
    \label{fig:noise_1d}
\end{centering}
\end{figure*}
\begin{figure*}
\begin{centering}
    \includegraphics[width=0.48\textwidth]{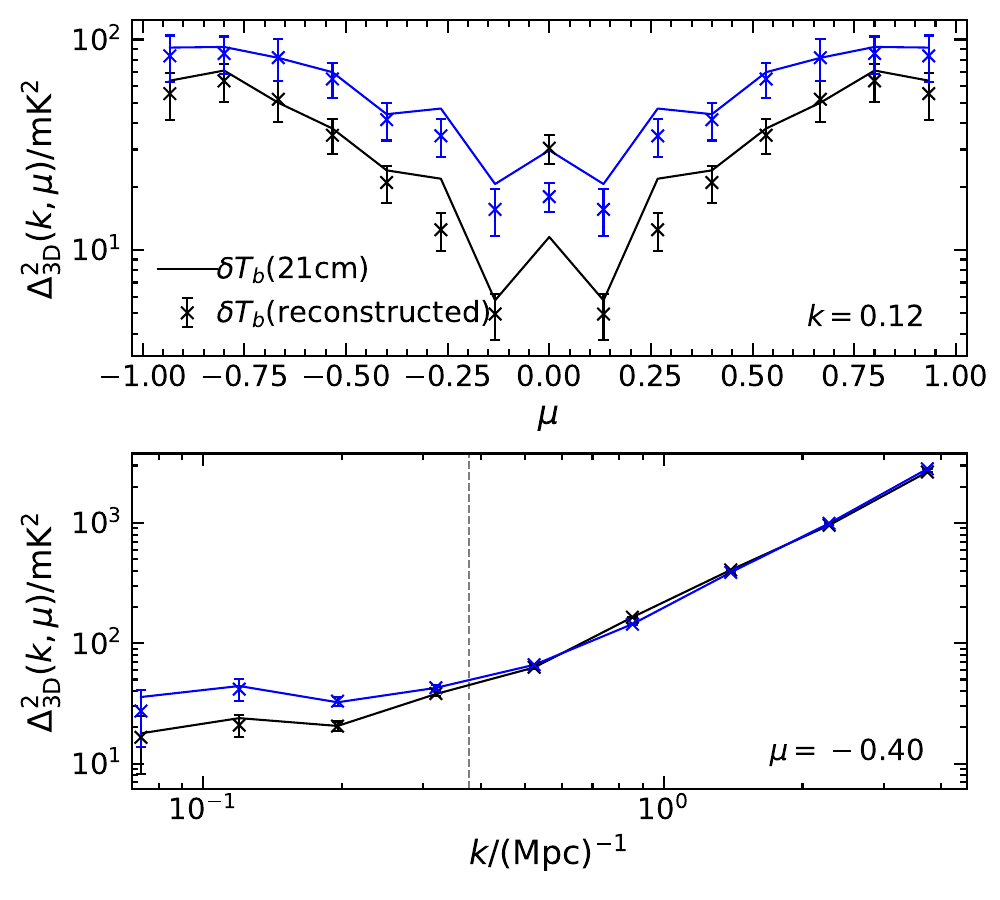}
    \includegraphics[width=0.48\textwidth]{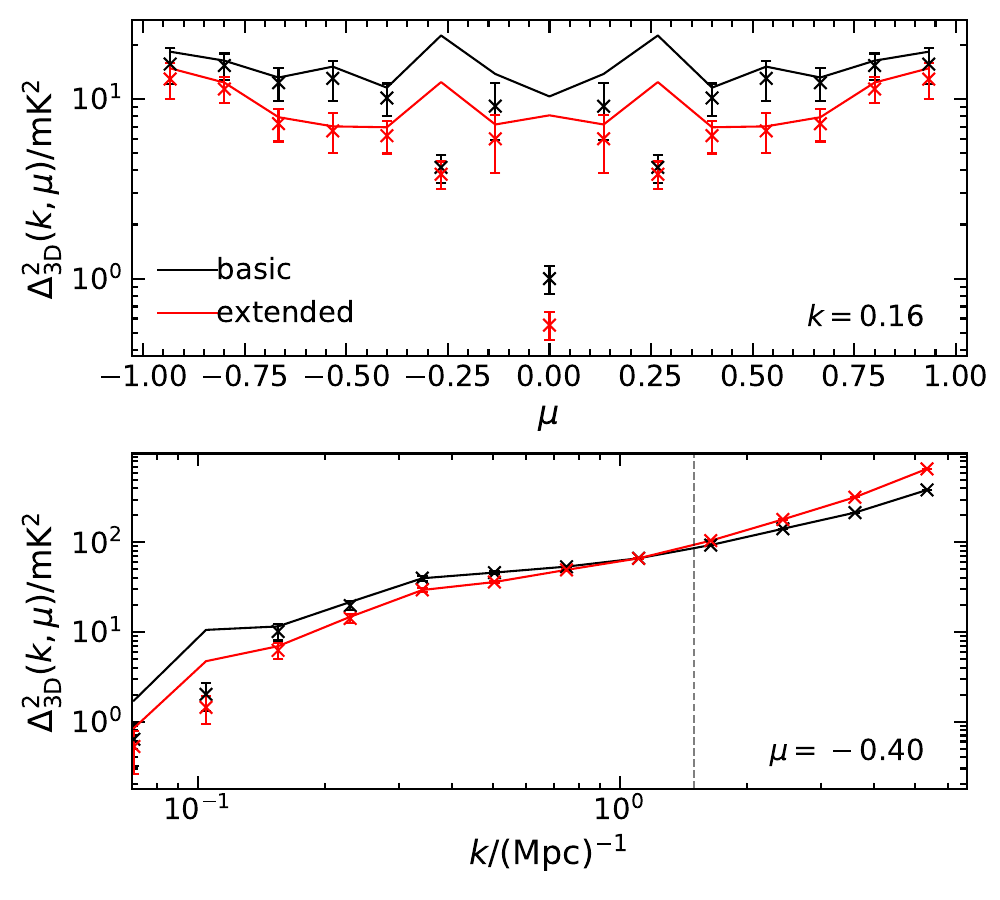}
    \vspace{-0.4cm}
    \caption{The 3D power spectrum decomposed into $\mu$ and $k$ for two 10 MHz lightcones at {\it{left:}} 80-90 MHz and {\it{right:}} 150-160 MHz, with lines as described as in Fig~\ref{fig:noise_1d}. The vertical dashed line in the bottom panel shows the $k$ at which the power of the noise for the SKAO-low AA4 surpasses the power of the signal as seen in Fig~\ref{fig:noise_1d}. The top panels highlight the potential to recover small $k$ scales at large $|\mu|$.}
    \label{fig:noise_mu}
\end{centering}
\end{figure*}

\begin{figure*}
\begin{centering}
\begin{subfigure}[b]{1.0\textwidth}
    \includegraphics[width=\textwidth]{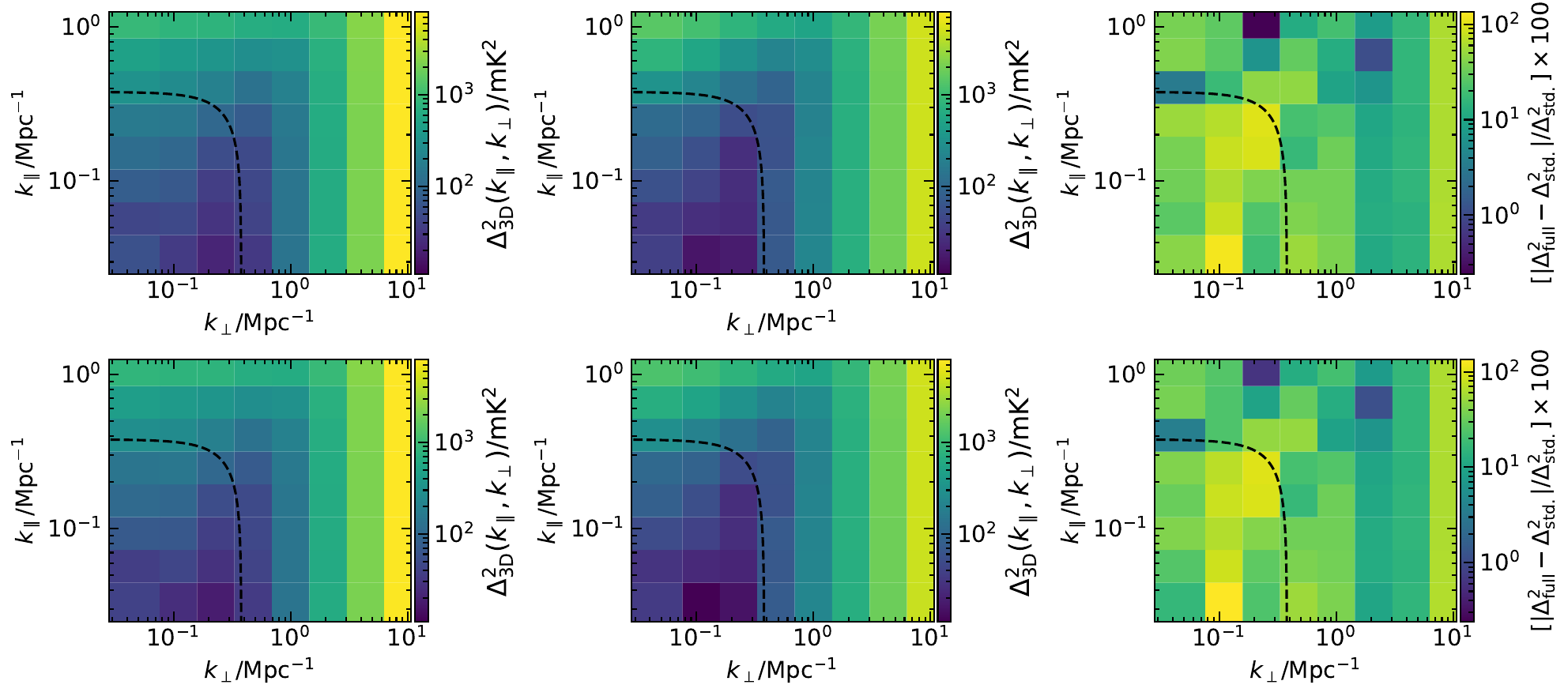}
    \caption{80 - 90 MHz}
    \end{subfigure}
\vfill
     \begin{subfigure}[b]{1.0\textwidth}
    \includegraphics[width=\textwidth]{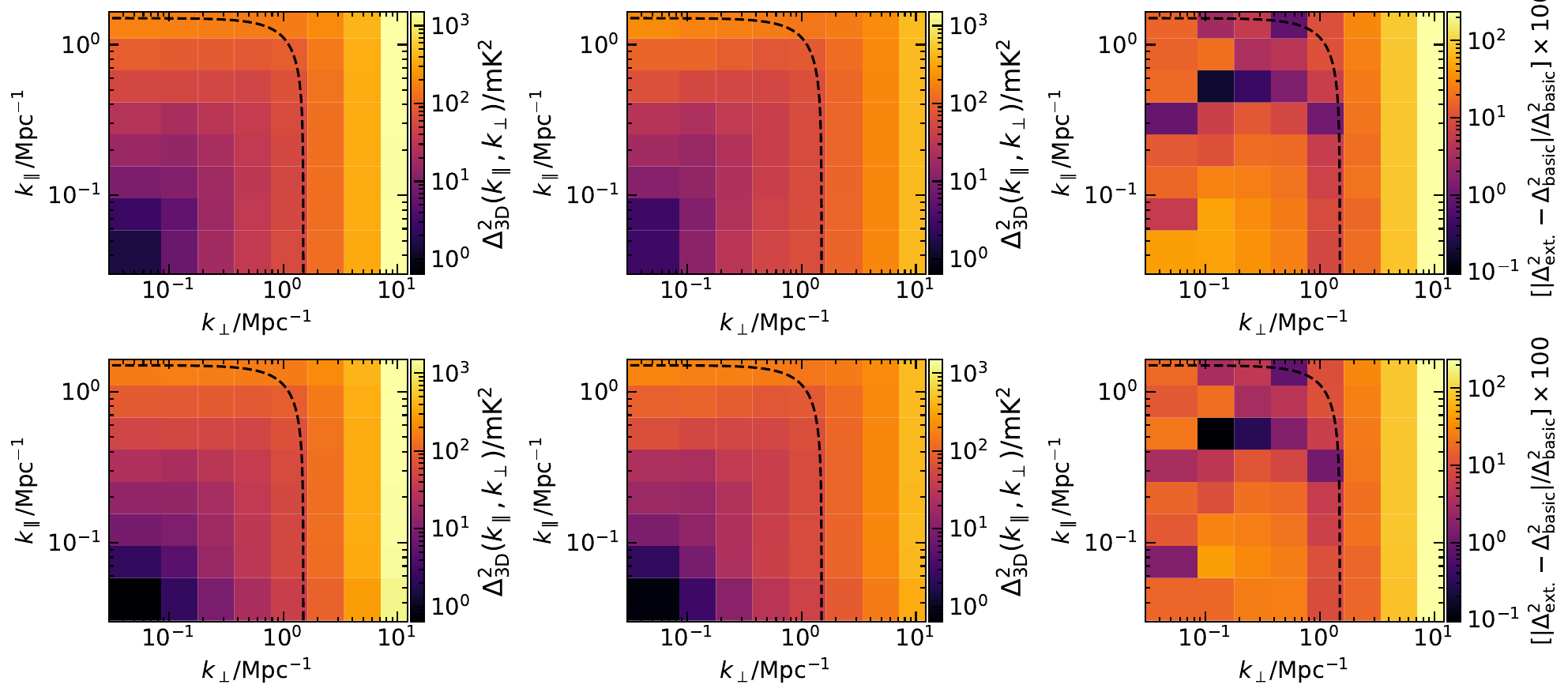}
    \caption{140 - 150 MHz}
    \end{subfigure}
    \vspace{-0.4cm}
    \caption{The 2D cylindrical averaged power spectra for a $10$ MHz lightcone, in the Cosmic Dawn ({\it{(a)}} 80-90 MHz) and the EoR ({\it{(b)}} 150-160 MHz). In each subfigure, the top row shows the power spectrum of the true $21$ cm signal and the bottom row shows the reconstructed signal. The columns left to right show our extended method, the basic method and the absolute difference between the two. We see small differences across the majority of $k$ scales. The k-scales with a power spectrum value below the sensitivity of a 1000-hour observation with the SKAO-low AA4 are shown to the right of the black dashed line, as taken from Fig.~\ref{fig:noise_1d}}
    \label{fig:Cyclindrical_ps_NOISE}

    \end{centering}
\end{figure*}
We now turn to investigate whether the differences seen between these two methods of creating a $21$ cm lightcone persist in a more observationally realistic scenario, and if the extended treatment of the peculiar velocity effect can aid in a detection of the 21cm signal. In this paper we investigate the addition and removal of diffuse radio foregrounds and SKAO-like instrumental noise. \cite{Jensen_RSD_2013} investigated the effects RSDs and the lightcone effect have when uncoupled in the presence of diffuse foregrounds and LOFAR-like noise, and found that the effects of RSDs are observable with $>1000 \,\rm hours$ of LOFAR time at scales of $k \approx 1 \,\rm Mpc^{-1}$ using a wavelet based foreground removal strategy.

The initial conditions were created on a $1200^3$ grid and smoothed onto a $600^3$ grid of length $200$ cMPc. We then created two lightcones for each method, one highlighting the Cosmic Dawn at $\nu_{0}=75- 95$ MHz ($z=17.9-14.0, \, x_{\rm HII}=0.02 - 0.00$) and one highlighting the middle of the EoR at $\nu_{0}=140- 180$ MHz ($z=9.1 - 6.9, \, x_{\rm HII}=0.14 -0.80$). The reionisation history used in this paper is shown in Fig.~\ref{fig:xHI}, in which we can see that our model is consistent with current ionisation fraction constraints. The frequency ranges for both lightcones used have been shown with vertical dashed lines (red for the EoR and blue for the Cosmic Dawn). All lightcones were made with a fixed field of view of $1 \degree$, an arcsecond resolution of $\frac{\theta}{N_{pixel}}=7''$, a frequency spacing of 0.1 MHz, and a coeval box spacing of $\Delta z = 0.1$. We refer to the different frequency ranges as the EoR (140 $-$ 180 MHz) and Cosmic Dawn (75 $-$ 95 MHz) lightcones.

The radio foregrounds added are diffuse galactic synchrotron and free-free emission. We have assumed that a calibration step has already taken place and that extragalactic foregrounds and bright point sources were adequately removed in this process. The foreground lightcones were made using the same resolution and frequency ranges as the $\delta T_{\rm b, 21 \rm cm }  $ lightcones and have been simulated using the methods outlined in \cite{Jelic_foregrounds_2008}. These foregrounds have a power approximately $100 \times$ larger than the $21$ cm signal simulated in this paper. The contaminated signal was made by adding the $21$ cm signal and diffuse foregrounds.

We have used version 2.0 of the Python package 21cmSense \citep{Murray_21cmSENSE_2024} to simulate the thermal noise level of the full SKAO-low area when observing for 1000 hours. 21cmSense uses the method outlined in \cite{Pober_sense_2013} and \cite{Pober_sense_2014} and summarised in \cite{Liu_Shaw_2020}. Note, no instrumental noise is added to the foreground cube itself, however we do calculate the noise power spectra in order to estimate the range of $k$ where any recovered 21 cm signal might be detectable. When calculating the sensitivity, we use the AA* and AA4 layouts of the SKAO low, which uses 307 and 512 stations \footnote{\url{https://www.skao.int/en/science-users/ska-tools/494/ska-staged-delivery-array-assemblies-and-subarrays}}. The sensitivity limit we calculate represents the theoretical power limit of the instrument. Although removing foregrounds down to this limit has not yet been achieved, due to excess noise features that are yet to be identified and fully removed. 

To model out the contaminating foregrounds, we have used the non-parametric foreground removal technique FastICA, outlined in \cite{Chapman_fastica_2012}. FastICA models the foregrounds through their non-Gaussianity and assumes that each component is statistically independent. The created model is then subtracted from the contaminated lightcone, creating a residual data cube of the $21$ cm signal and any foreground leakage. In this paper, we focus on the ability to see a difference between the two methods after foreground removal rather than comparing different foreground removal methods. Therefore, we consistently use a four-component fit of FastICA, which has been fitted over the entire lightcone for all the analysis presented in this paper.

To fully quantify the differences seen between these two methods, we look at three different power spectra. The first is the 3D power spectrum decomposed in $k$ defined as $\Delta^2_{3D}=\frac{k^3P(k)}{2\pi^2V}$, where $V$ is the volume of the lightcone to be analysed, $P(k)=\langle\delta(k)\delta^*(k)\rangle$ and $\delta(k)$ is the standard Fourier transform. The average for different $k$ scales is calculated for different radii of spherical shells in $k$-space. The second is the 3D power spectrum decomposed in $\mu$, where $\mu$ denotes the angle with the line-of-sight ($\mu=k_{\parallel}/|k|$), for each $k$ mode. In this presentation of a power spectrum, an isotropic $k$ mode would cause $P_\mu (k,\mu)$ to be a flat line. Using $P_\mu (k,\mu)$ enables us to highlight different $k$ modes individually and hence investigate the differences between the two methods more thoroughly. We also look at the cylindrical power spectra, as it highlights the line-of-sight effects. This is calculated through applying a 3D Fourier transform on the data and then binning the voxels by perpendicular ($k_{\perp}$) and parallel ($k_{\parallel}$) Fourier scales. For all our analysis, we calculate each power spectrum over 10 MHz sections of the residual lightcones, and errors on the power spectra are shown by $\frac{1}{\sqrt{n}}$ where $n$ is the number of modes in each bin. We note that the efforts to detect this signal lie mainly with the cylindrical power spectrum, we include 1D and $\mu$ power spectra for theoretical insight.

The 3D power spectrum decomposed in $k$ is shown in Fig.~\ref{fig:noise_1d}. We see the thermal noise sensitivity of the SKAO (grey shaded region) alongside the true (solid lines) and reconstructed signal (crosses) for both methods. The basic method is shown in black, and our extended method is shown in red for the EoR and blue for the Cosmic Dawn. The bottom panels show the power spectrum ratio of the extended method to the basic method. We see that the $<300\%$ ($<50\%$) enhancement at the largest $k$ scale investigated during the EoR (Cosmic Dawn) is recoverable after the addition and removal of diffuse foregrounds. Therefore, if using an instrument with adequate sensitivity, this boost at large $k$ will aid the detectability of the small scale structures. This figure also shows that the suppression at scales $k<1.0$ Mpc during the EoR is not well recovered, due to the diffuse foregrounds dominating at these scales. From examining the noise sensitivity of the SKAO, we observe that at scales of $k > 1.50$ Mpc ($k > 0.38$ Mpc) in the EoR (Cosmic Dawn), the power of the noise sensitivity levels exceeds that of the 21cm signal, making any observation of the signal challenging. We note that the choice of array layout has little effect on the observability of differences between the two methods, and we only show the AA4 sensitivity limit in all further plots to aid in readability.

The 3D power spectrum decomposed in $\mu$ is shown in Fig.~\ref{fig:noise_mu} for the basic (black) and extended (red/blue) method. The true $21\rm cm$ signal is shown in the solid line, and the reconstructed signal is shown as crosses. The dashed line in the bottom panel reflects the scale at which the sensitivity of the instrument surpasses that of the $21$ cm signal, as shown in Fig.~\ref{fig:noise_1d}. We note that although the sensitivity of the SKAO-low does not allow recovery of the enhancement seen at large $k$, this is in the regime where the foregrounds do not dominate. We do see a small enhancement (suppression) in the $k$ scales, which are not dominated by the sensitivity of the interferometer during the Cosmic Dawn (EoR). We have highlighted the non-noise dominated $k$ modes in the upper panel to demonstrate that differences between the methods persist across all $\mu$ scales, and that with good foreground removal, it is possible to distinguish between them. Deconstructing the power spectrum into different $k$ and $\mu$ modes has also made the differences between the two methods more visible.

The cylindrical power spectrum is shown in Fig.~\ref{fig:Cyclindrical_ps_NOISE}, for the Cosmic Dawn (a) and EoR lightcone (b). The top rows of each sub-figure contain the true $21$ cm signal above their corresponding reconstructed signal. The columns from left to right show the extended method, basic method and the percentage of absolute difference between the two, $(|\Delta^2_{\rm full}-\Delta^2_{\rm std}|/\Delta^2_{\rm std})\times 100$. A black dashed line has been drawn to show the $k$ scales where the sensitivity of the telescope exceeds that of the $21$ cm signal. It can be seen that although the noise contaminates the largest $k$ scales where an enhancement of $>100\%$ is seen for the extended method relative to the basic method in the EoR, we do see differences of $\approx 10\%$ between the two methods over the recoverable scales. Also, we show that robust inclusion of peculiar velocity effects leads to observationally significant differences of $>100$\% that are recoverable in the presence of diffuse foregrounds. The ability to recover the differences between the two methods looks more promising in the Cosmic Dawn, where we see some of the up to $100 \%$ differences between the two methods to be observationally significant.

\section{Conclusions}

We have presented a new method that treats the light-cone effect and redshift-space distortions in a fully coupled framework. By applying both effects simultaneously, the resulting 21-cm maps more accurately reflect the true evolution of the signal along the line of sight, capturing multiple signal components within a single observed voxel.
 These effects have often been investigated individually or in a semi-coupled fashion \citep[e.g.][]{Pramanick_resolution_lightcone_2023, Ross_RSD_2021, Mondal_lightcone_effect_2018, Jensen_RSD_2013, Datta_lightcone_2012} but they were first looked at in conjunction with each other in CS19. When fully incorporating these effects, we see a change in all types of power spectra investigated. Most notably, the cylindrical power spectra. Due to removing common assumptions in $21$ cm semi-numeric codes, our extended method allows for more than one $21$ cm contribution to each lightcone voxel, and photons to be red/blue shifted from different slices and boxes of redshift snapshots. We have investigated how each of these effects contributes to differences between the two methods, and the required spatial, frequency, and redshift resolution needed to make a converged lightcone regardless of the method used. Finally, we looked at the observability of the differences in the power spectra of both methods. To do this, we have added and removed diffuse radio foregrounds and simulated the sensitivity of the SKAO-Low for a 1000-hour observation using the AA* and AA4 layouts. The foreground removal method was kept consistent throughout all analyses to allow for a fair comparison. Our main findings were as follows:

\begin{itemize}
    \item On average 8\% of a coeval cell contributes to a lightcone voxel for the extended method, compared to 100\% for the basic method. These partial $21$ cm contributions have been seen to be red/blue-shifted from a $\pm 0.5$ MHz range from the expected emitted frequency. The brightest voxels in the lightcone consisted of 120\% of a coeval cell, showing that the extended method allows for much brighter 21cm lightcone voxels, which come from only 20\% more coeval space. These effects scale with coeval box and lightcone resolution quality. \\
    
    \item It was seen that redshift snapshot spacing was the most important resolution choice when constraining the end stages of reionisation. Although we note that a balance must be struck between the $\Delta z$ spacing of redshift snapshot boxes and computational costs. \\ 

    \item We showed that the differences in the power spectrum of the $21$ cm lightcones seen when using the basic and extended method are recoverable after the addition and removal of diffuse radio foregrounds. These differences were best highlighted when using the cylindrical power spectrum for both the cosmic dawn and epoch of reionisation, which is the most promising power spectrum method for a detection of the 21cm brightness temperature. The enhancement seen when using the extended method in the cosmic dawn will improve the detectability of the power spectra of the $21$ cm signal.\\

    \item The sensitivity of an 1000 hour SKAO-low observation using the AA4 layout has been shown to surpass the level of the power spectra of the $21$ cm signal at $k > 1.5$ Mpc ($k > 0.38$ Mpc) in the EoR (Cosmic Dawn) in the 3D power spectrum in k. However, in the cosmic dawn, the biggest differences between the two methods are still observable in the $\mu$ or cylindrical power spectra, and hence the presence of thermal noise is not an insurmountable obstacable. We leave incorporating more complex instrumental effects on the data cube and inclusion of extragalactic foregrounds to future work. It would also be of interest to repeat this analysis with more exotic models of reionisation to see the effect of reionisation history on the two methods. We note that although our more robust method of simulating lightcones of $21$ cm brightness temperature is more computationally intensive, it is an important and potentially observable distinction to make. 

\end{itemize}
\section*{Acknowledgements}
JF acknowledges the support of a studentship from the Science \& Technology Facilities Council. EC acknowledges the support of a Royal Society Dorothy Hodgkin Fellowship and a STFC consolidated grant ST/X000982/1. The authors would like to acknowledge Mario G. Santos and Luke Conaboy for their useful discussion. We also thank the referee for a constructive report that helped to improve this manuscript. 

\section*{Data Availability}
All data and analysis code used in this work are available from the first author on reasonable request.



\bibliographystyle{mnras}
\bibliography{refs}





\bsp	
\label{lastpage}
\end{document}